\documentclass[aps,prb,twocolumn,superscriptaddress,showpacs]{revtex4}
\usepackage{graphicx}
\usepackage{longtable}
\usepackage{array}
\usepackage{braket}
\newcolumntype{L}{>{\centering\arraybackslash}m{2cm}}
\newcolumntype{R}{>{\centering\arraybackslash}m{1.5cm}}
\newcolumntype{K}{>{\centering\arraybackslash}m{1.3cm}}

\def \NO{Nd$_2$O$_3$}

\begin{document}


\title{Physical properties of the trigonal binary compound \NO\footnote{This manuscript has been authored by UT-Battelle, LLC under Contract No. DE-AC05-00OR22725 with the U.S. Department of Energy.  The United States Government retains and the publisher, by accepting the article for publication, acknowledges that the United States Government retains a non-exclusive, paid-up, irrevocable, world-wide license to publish or reproduce the published form of this manuscript, or allow others to do so, for United States Government purposes.  The Department of Energy will provide public access to these results of federally sponsored research in accordance with the DOE Public Access Plan (http://energy.gov/downloads/doe-public-access-plan).}}

\author{G. Sala}
\email{salag@ornl.gov}
\affiliation{Neutron Scattering Division, Oak Ridge National Laboratory, Oak Ridge, Tennessee 37831, USA}

\author{M. B. Stone}
\email{stonemb@ornl.gov}
\affiliation{Neutron Scattering Division, Oak Ridge National Laboratory, Oak Ridge, Tennessee 37831, USA}

\author{B. K. Rai}
\affiliation{Materials Science \& Technology Division, Oak Ridge National Laboratory, Oak Ridge, TN 37831, USA}

\author{A. F. May}
\affiliation{Materials Science \& Technology Division, Oak Ridge National Laboratory, Oak Ridge, TN 37831, USA}

\author{C. R. Dela Cruz}
\affiliation{Neutron Scattering Division, Oak Ridge National Laboratory, Oak Ridge, Tennessee 37831, USA}

\author{H. Suriya Arachchige}
\affiliation{Department of Physics \& Astronomy, University of Tennessee, Knoxville, TN 37996, USA}
\affiliation{Neutron Scattering Division, Oak Ridge National Laboratory, Oak Ridge, TN 37831, USA}

\author{G. Ehlers}
\affiliation{Neutron Technologies Division, Oak Ridge National Laboratory, Oak Ridge, TN 37831, USA}

\author{V. R. Fanelli}
\affiliation{Neutron Scattering Division, Oak Ridge National Laboratory, Oak Ridge, Tennessee 37831, USA}

\author{V. O. Garlea}
\affiliation{Neutron Scattering Division, Oak Ridge National Laboratory, Oak Ridge, Tennessee 37831, USA}

\author{M. D. Lumsden}
\affiliation{Neutron Scattering Division, Oak Ridge National Laboratory, Oak Ridge, Tennessee 37831, USA}

\author{D. Mandrus}
\affiliation{Department of Physics \& Astronomy, University of Tennessee, Knoxville, TN 37996, USA}
\affiliation{Materials Science \& Technology Division, Oak Ridge National Laboratory, Oak Ridge, TN 37831, USA}
\affiliation{Department of Material Science \& Engineering, University of Tennessee, Knoxville, TN 37996, USA}

\author{A. D. Christianson}
\email{christiansad@ornl.gov}
\affiliation{Materials Science \& Technology Division, Oak Ridge National Laboratory, Oak Ridge, TN 37831, USA}

\date{\today}

\begin{abstract}
We have studied the physical properties of \NO{} with neutron diffraction, inelastic neutron scattering, heat capacity, and magnetic susceptibility measurements. \NO{} crystallizes in a trigonal structure, with Nd$^{3+}$ ions surrounded by cages of 7 oxygen anions. The crystal field spectrum consists of four 
excitations spanning the energy range 3-60 meV. The refined eigenfunctions indicate XY-spins in the $ab$ plane. The Curie-Weiss temperature of $\theta_{CW}=-23.7(1)$ K was determined from magnetic susceptibility measurements.  Heat capacity measurements show a sharp peak at 550 mK and a broader feature centered near 1.5 K. Neutron diffraction measurements show that the 550 mK transition corresponds to long-range anti-ferromagnetic order implying a frustration index of $\theta_{CW}/T_N\approx43$.  These results indicate that \NO{} is a structurally and chemically simple model system for frustration caused by competing interactions with moments with predominate XY anisotropy.
\end{abstract}




\maketitle



\section{Introduction}

Nd$^{3+}$ based compounds have recently been a subject of renewed interest in relation to the so-called ``moment fragmentation" mechanism~\cite{Ludovic}. 
This puzzling phenomenon was recently put forward in the context of spin ice, in which the magnetic moment can fragment, resulting in a dual ground state (GS) consisting 
of a fluctuating spin liquid, a so-called Coulomb phase, superimposed on a ``magnetic monopole" crystal. Experimentally this fragmentation can be realized in e.g. $\rm Nd_2Zr_2O_7$~\cite{Lake1,Benton} 
and $\rm Nd_2Hf_2O_7$~\cite{Lake2} where it manifests in neutron diffraction measurements as the superposition of magnetic Bragg peaks, characteristic of the ordered phase, and a pinch point pattern, 
characteristic of the Coulomb phase. These results highlighted the relevance of the fragmentation concept to describe the physics of systems that are simultaneously
ordered and fluctuating, and it opens the possibility to study new exciting phenomena in frustrated systems. 

Additional examples of the aforementioned behavior are sought by the community.  Beyond this, simple examples which may help establish trends in magnetic properties of such materials are important.  In this regard, \NO{} appears potentially interesting.   Indeed, previous characterizations of \NO~ found no magnetic order above 4 K~\cite{Hacker, Justice}, with Curie-Weiss temperatures in the range $-20 \leq \Theta_{CW} \leq -32$ K suggesting that \NO~may be a structurally and chemically simple model system with significant frustration and competing interactions. Additionally, \NO~is a potentially important magnetic impurity phase in the context of other more complex Nd-based materials and a thorough characterization of the low temperature physical properties is thus important.

\begin{figure}[t]
\includegraphics[width=0.85\columnwidth]
                {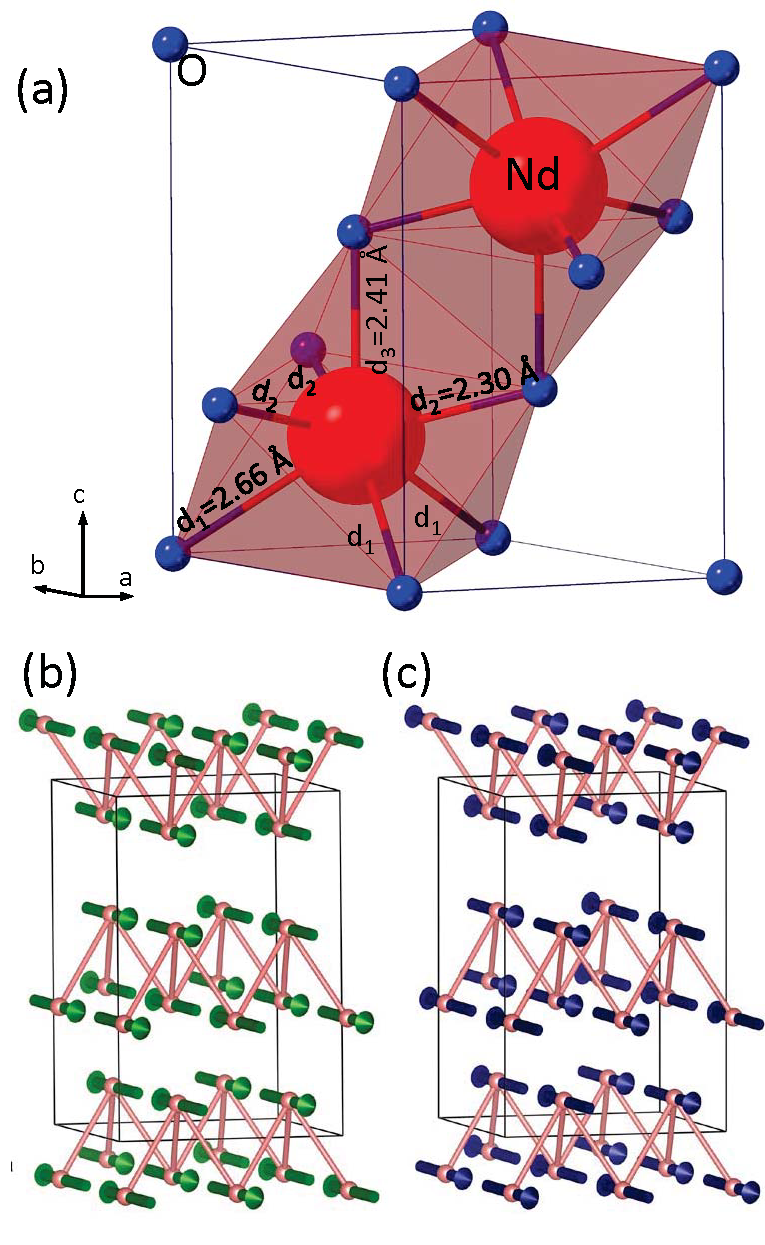}
\caption{
\label{fig1}
(a) Crystal structure of \NO with red spheres representing Nd sites and blue spheres representing oxygen sites.    \NO~crystallizes in the trigonal centrosymmetric space group $P\bar{3}m1$ (164). The environment surrounding the Nd$^{3+}$ sites consists of 7 oxygen ions located at the vertex of a distorted cube along one diagonal.
The $C_3$ axis coincides with the c-axis, while one of the mirror planes lies perpendicular to the ab plane.   (b) and (c) Refined magnetic structures of \NO{} showing only the magnetic moments on the Nd sites. The Rietveld refinement at 280 mK indicates two potential magnetic structures, both with antiferromagnetically aligned moments in the ab-plane and either (b)($--++$)  or (c)($+--+$) stacking along c-axis.}
\end{figure}

As shown in Fig.~\ref{fig1}a, \NO~crystallizes in the trigonal centrosymmetric space group $P\bar{3}m1$ (164); this space group is in agreement with the classification made
by Pauling~\cite{Pauling}, Steven~\cite{Stevie}, Faucher~\cite{Faucher} and, more recently, by Gruber~\cite{Gruber} which also predicts the correct symmetry of the Raman modes. Other authors assigned
to this compound the acentric space group $P321$~\cite{Zachariasen}, or $P6_3/mmc$~\cite{Muller}. As we describe in greater detail below, both of these options have been ruled out by 
our refinement at $T = 2$ K which gives the best agreement with the data set for the trigonal $P\bar{3}m1$ space group, having
a = b = 3.83(1) \AA, c = 6.00(3) \AA, $\alpha$ = $\beta = $ 90$^{\circ}$ and $\gamma = $ 120$^{\circ}$ in accord with previous measurements~\cite{Pauling,Stevie,Faucher,Gruber}.

The environment surrounding Nd$^{3+}$, shown in Fig.~\ref{fig1}(a), has a $C_{3v}$ symmetry, consisting of 7 oxygen ions located at the vertex of a distorted cube along one diagonal. The $C_3$ axis coincides with the c-axis, while one of the mirror plane lies perpendicular to the ab plane. Nd ions are arranged on triangular nets stacked along the c-axis. The two shortest Nd-Nd bonds (3.69 and 3.77 \AA) are along the c-axis.  The shortest Nd-Nd distance in the plane is somewhat longer than those out of the plane at 3.83 \AA. 
These three similar bonds distances likely result in competing interactions in addition to the geometric frustration inherent to the triangular nets of Nd.

The goal of the present study is to characterize the magnetic properties of \NO, particularly at lower temperatures than have been previously reported. Fits to the magnetic susceptibility with a Curie-Weiss law find an effective moment of 3.64(1) $\mu_B$/ion and a $\theta_{CW} = -23.7(1)$ K.  Inelastic neutron scattering measurements reveal a spectrum of 4 crystal field excitations from the ground state doublet to excited states. We analyze this spectrum to determine the crystal field Hamiltonian.  The ground state eigenfunction extracted from this analysis indicates moments with XY anisotropy.   Low temperature heat capacity measurements show a broad peak centered at 1.5 K and a sharp peak at 550 mK not previously reported. Neutron diffraction measurements further reveal that the sharp peak in the heat capacity data corresponds to the onset of long range magnetic order at 550 mK with a ordering wave vector $\vec{k}$ = (1/2 0 1/2).  The magnetic structure determined from the diffraction data is characterized by moments lying in the trigonal basal plane with an ordered moment size of 1.87(10)$\mu_B$ (see Figs.~\ref{fig1}(b,c)). Together the results presented here indicate that \NO~is a chemically simple example of a frustrated magnet with moments with a strong XY anisotropy on a centrosymmetric lattice.


\section{
\label{sec: Exp}
Experimental Details}

\NO was obtained from Alfa Aesar (99.997\%) and dried in air at 1050$^{\circ}$C.  Using the same source material, pellets of diameter $\approx$ 8\,mm were prepared by cold-pressing dried powder and sintering in air at 1250$^{\circ}$C for 12\,h.  These samples were characterized by means of a Quantum Design Physical Property Measurement System and a Magnetic Property Measurement System.  Specific heat data were collected using an $^3$He insert in the temperature range
$0.4 \leq T \leq 10$ K with an applied magnetic field $H$ of zero and 50\,kOe. The magnetization $M$ was measured upon cooling in $H$ = 10\,kOe, and the results are reported as the magnetic susceptibility $\chi$=$M$/$H$.

The crystal field excitations have been studied using the SEQUOIA spectrometer at the Spallation Neutron Source at Oak Ridge National Laboratory~\cite{SEQref}. Approximately 5\,g of polycrystalline \NO~was loaded into a cylindrical Al can and sealed under helium exchange gas. The sample and an equivalent empty can for background subtraction~\cite{aluminum} were mounted in a sample carousel. Measurements were performed at 5 K and 50 K, with incident energies, $E_is$, of 25, 75 and 130 meV. Higher energy resolution measurements were performed on the same sample with the Cold Neutron Chopper Spectrometer (CNCS)~\cite{CNCSref}. For these measurements the sample was cooled to 2 K in a cryo-magnet and  measured with $E_i=$ 6.5 meV at 0 
and 50 kOe. Unless otherwise noted, all inelastic measurements presented here have had the measured backgrounds subtracted.  The collected data sets from SEQUOIA and CNCS have been analyzed using the software MANTID~\cite{mantidplot} and Dave~\cite{dave}.

The magnetic structure has been studied through neutron powder diffraction measurements with the HB-2A powder diffractometer at the High Flux Isotope Reactor (HFIR)~\cite{HBref}.  Sintered pellets were stacked to achieve a sample of cylindrical shape 0.4 mm radius and 1.5 cm length. The pellets were loaded into a Cu cylindrical can to maximize thermal contact, and sealed with a He exchange gas. Measurements with $\lambda$ = 2.54 \AA~were conducted in zero applied field at 2 K, 1 K and 280 mK using a cryostat with a $^{3}$He insert.  The magnetic structure refinement was performed independently with the software Full-Prof~\cite{Fprof} and JANA2006~\cite{Jana}.


\begin{figure}[t]
\includegraphics[width=1.0\columnwidth]
                {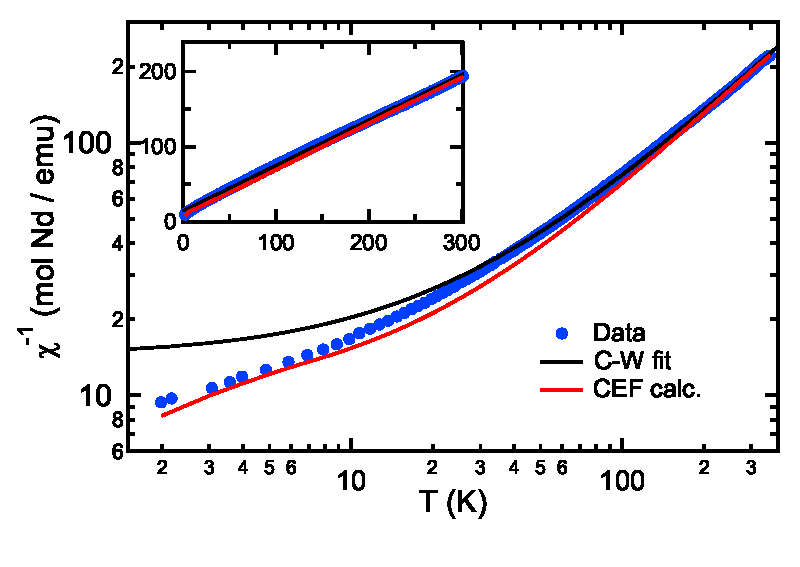}
\caption{
\label{fig: susc}
\NO~susceptibility collected over the temperature range $2\leq$ T $\leq$300 K (blue dots). The Curie-Weiss fit was performed above 50\,K and yielded $\theta_{CW} = -23.7(1)$ K and an effective magnetic moment of 3.64(1) $\mu_B$/ion. The red line shows the calculated crystal field 
susceptibility based on our refinement in LS-coupling, we ascribe the small deviations below $\approx 25$ K to Nd-Nd spin correlations.  Inset shows the same data on a linear-linear scale.}
\end{figure}

\section{Crystal Field Analysis}

Inelastic neutron scattering is an efficient probe of crystalline electric field (CEF) excitations.  Characterization of the crystal field levels allows the nature of the magnetic ground state of a system to be determined with reasonable accuracy.  The crystal field analysis performed here follows the formalism described by Wybourne~\cite{Wybourne,Judd1,Judd2}; given the local structure of \NO~drawn in Fig.~\ref{fig1}(a), and the $C_{3v}$ site symmetry of this compound, the crystal field Hamiltonian consists of $6$ parameters~\cite{Walter}. Prather's convention~\cite{Prather} for the minimal number of crystal field parameters was achieved by rotating the environment by $\pi/6$ along the three fold axis.  Finally, within LS coupling, Hund's rules state that, for a $4f^3$ ion, the quantum numbers are $L=6$ and $S=3/2$, thus $J = \| L - S\| = 9/2~$~\cite{Kittel}.  Therefore the crystal field Hamiltonian can be written as:
\begin{eqnarray}
H = B_2^0\hat{C}_2^0 + B_4^0\hat{C}_4^0 + B_4^3(\hat{C}_4^3-\hat{C}_4^{-3}) + B_6^0\hat{C}_6^0 
\nonumber
\\
+ B_6^3(\hat{C}_6^3-\hat{C}_6^{-3}) + B_6^6(\hat{C}_6^6+\hat{C}_6^{-6})
\label{eq: 1}
\end{eqnarray}
where $B_n^m$ are the crystal field parameters and $\hat{C}_n^m$ are spherical tensor operators~\cite{Racah1,Racah2,Racah3,Racah4}.
Nd$^{3+}$ is a Kramers ion, so we expect to have at least doubly degenerate CEF levels as stated in the theorem~\cite{Kramer}. The effect on the Hamiltonian of a Zeeman term, 
will result in breaking time reversal symmetry, splitting the degeneracy of the levels from doublets to singlets. As a first approximation this splitting will be proportional 
to the strength of the field times the magnetic moment of the ion thus, at small fields, this term will broaden the crystal field levels and decrease their intensity.

Once Eq.~\ref{eq: 1} is diagonalized, the unpolarized neutron partial differential magnetic cross-section can be written within the dipole approximation as~\cite{squires}:
\begin{eqnarray}
\frac{d^2\sigma}{d\Omega dE'} = C\frac{k_f}{k_i}f^{2}(|Q|)S(|Q|,\omega)
\end{eqnarray}
where $\Omega$ is the scattered solid angle, $\frac{k_f}{k_i}$ the ratio of the scattered and incident momentum of the neutron, $C$ is a constant, $|Q|$ is the magnitude of the wave-vector transfer, and $f(|Q|)$ is the
magnetic form factor. The scattering function $S(|Q|,\hbar \omega)$ gives the relative scattered intensity due to transitions between different CEF levels. At constant temperature, $\beta = 1/k_BT$:
\begin{equation}
S(|Q|,\hbar\omega) = 
\sum_{i,i'}\frac{(\sum_{\alpha}  |\langle i {| J_{\alpha} | i'\rangle |}^2) \mathrm{e}^{-\beta E_{i}} }{\sum_j \mathrm{e}^{-\beta E_{j}}} F(\Delta E + \hbar \omega,\Gamma_{i,i'})
\label{eq: 2}
\end{equation}
where $\alpha = x,y,z$, $\Delta E = E_{i} - E_{i'}$, and $F(\Delta E + \hbar \omega, \Gamma_{i,i'})$ is a Lorentzian function with halfwidth $\Gamma$ to parameterize the lineshape of the transitions between the CEF levels $i \rightarrow i'$. We calculate the scattering function using this formalism.  The calculation is compared with experimental data set, and the procedure is iterated varying the crystal field parameters to minimize the $\chi^2$ difference.

\begin{figure*}
\includegraphics[width=1.75\columnwidth]
                {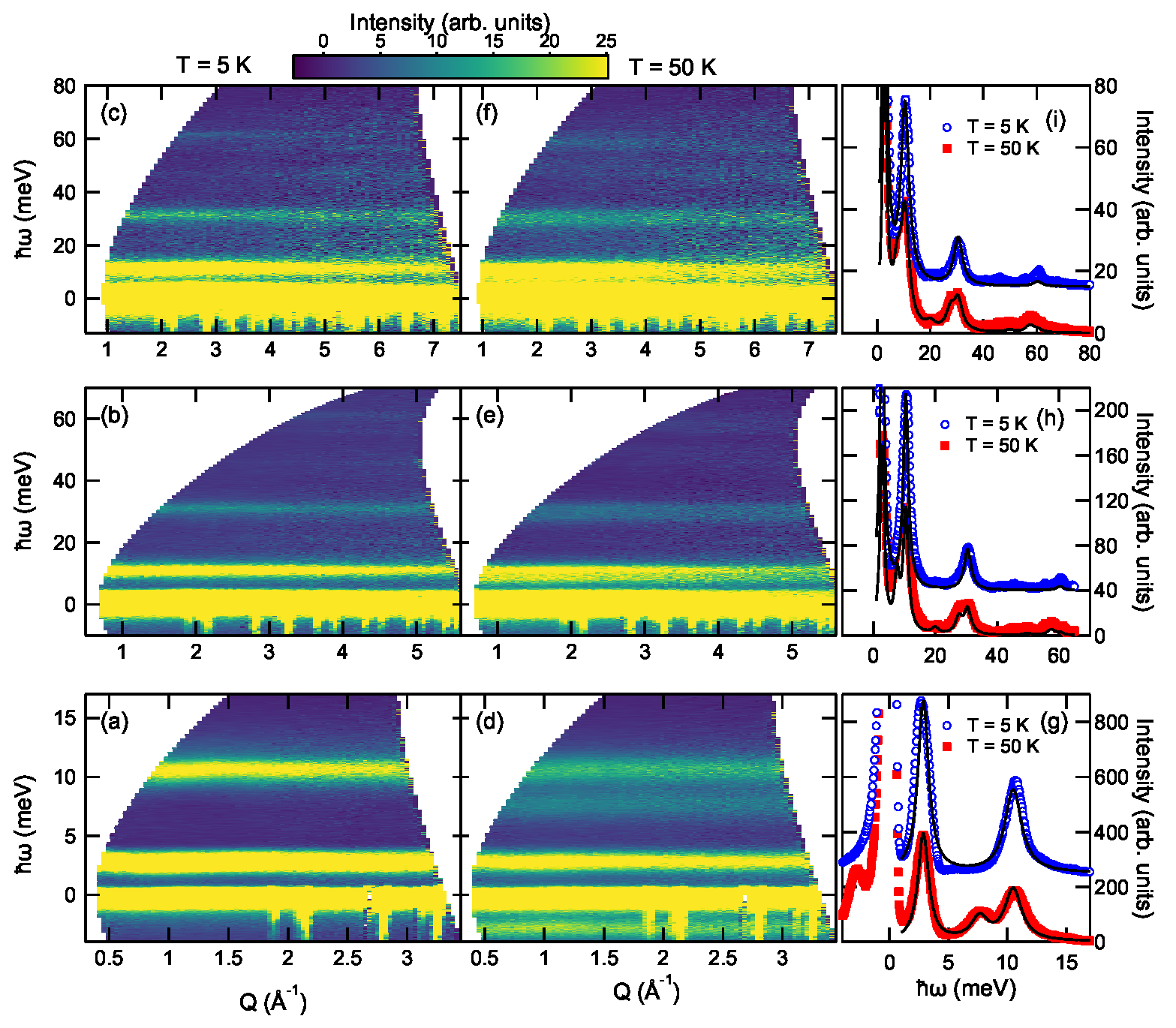}
\caption{
\label{fig: cf}
Inelastic neutron scattering spectra of \NO~measured at $T=5$~K (a,b,c) and $T=50$~K (d,e,f) using the SEQUOIA spectrometer. In (c) and (f) $E_i$= 130 meV; in (b) and (e)$E_i$= 75 meV; and in (a) and (d) $E_i$= 25 meV.  All slices have been plotted on the same intensity scale given at the top of the figure.  The feature at $\approx$ 58 meV is likely an oxygen phonon mode. As described in the text, the crystal field transitions are located at $2.86(3)$, $10.49(2)$, $30.46(3)$ and $60.29(4)$ meV.  (g), (h), and (i) show cuts of the spectra measured at $T=5$~K and $T=50$~K integrated over $0.5~$\AA$^{-1} < |\vec{Q}| < 3 ~$\AA$^{-1}$, $1~$\AA$^{-1} < |\vec{Q}| < 4.5 $\AA$^{-1}$, and $1~$\AA$^{-1} < |\vec{Q}| < 4.5$ \AA$^{-1}$ respectively. The solid lines in (g), (h), and (i) are the result of fits of the crystal field Hamiltonian model to the data as described in the text. The $T=5$~K data in panels (g)-(i) are offset by 250, 40 and 15 units respectively for clarity.}
\end{figure*}

A potential difficulty with the analysis described above is that the spectrum of Nd has higher J-multiplet levels which can be close to the GS one.  This implies that the eigenfunctions of the GS could result in a linear combination of different J-multiplets.  In order to investigate this possibility, we repeated our crystal field analysis in an intermediate coupling regime, including the first 12 J-multiplets up to $2$ eV. This analysis highlighted that there is indeed an effect due to the close proximity of the J = 9/2 and J = 11/2 multiplets but,
as shown in Tab.~\ref{tab: 3}, the overall GS eigenfunctions do not change appreciably from the original LS-approximation. Thus, our discussion of the physical properties of \NO{} in the remainder of the paper uses a description appropriate for a J = 9/2 multiplet rather than the more general case.


\section{Results and Discussion}

\subsection{Magnetic Susceptibility}

The magnetic susceptibility of \NO{} is shown in Fig.~\ref{fig: susc}.  A Curie-Weiss fit (black line) yields $\theta_{CW} = -23.7(1)$ K and an effective magnetic moment of 3.64(1)$\mu_B$/ion.  The  Curie-Weiss model $\chi$ = $C/(T-\theta_{CW})$ was utilized to fit data above 50\,K, where the Curie constant $C$ is related to the effective moment and $\theta_{CW}$ is the Weiss temperature.  The value of effective moment is comparable to the free ion value of 3.62 $\mu_B$~\cite{ashcroft}. At low temperatures, below $\sim$25 K, the Curie-Weiss fit deviates from the data as a consequence of the thermal depopulation of crystal field levels and the growing importance of magnetic correlations which ultimately drive the formation of a long-range magnetically-ordered state below 550 mK (see Sec. \ref{lro}).  The fitted curve in Fig. \ref{fig: susc} from the crystal field model (see Sec. \ref{cfs}) also deviates from the susceptibility data at a similar temperature, but in contrast to the CW model over estimates the susceptibility (and thus underestimates $\chi^{-1}$) presumably due to the importance of magnetic correlations at low temperatures. 


\subsection{Crystal Field Spectrum}\label{cfs}

\begin{figure}[t]
\includegraphics[width=0.95\columnwidth]
            {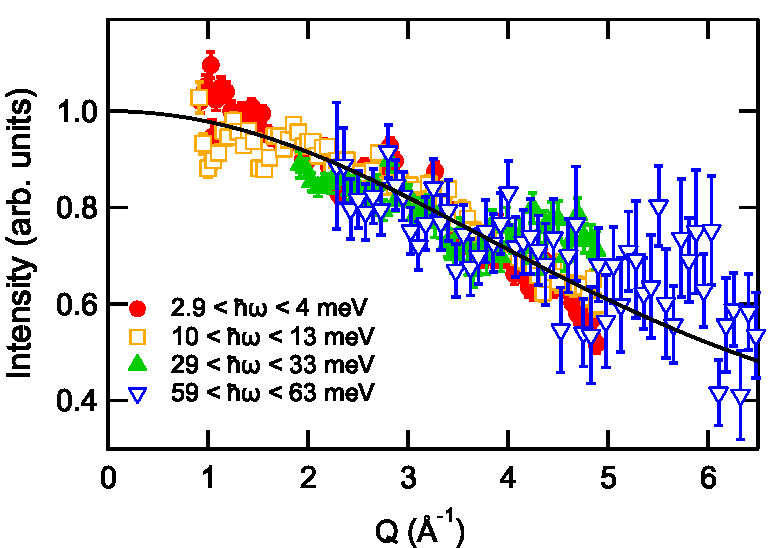}
\caption{
\label{fig:IvsQ}
Normalized scattering intensity as a function of $|Q|$ for \NO~measured using the SEQUOIA instrument as described in the text. Constant energy scans were integrated over energies as indicated in the legend of the figure. The solid line is the square of the Nd$^{3+}$ magnetic form factor.}
\end{figure}

Figure~\ref{fig: cf} shows an overview of the measured spectra collected at SEQUOIA as a function of energy transfer, $\hbar\omega$, and wave-vector transfer, $Q$, at $T=5$ and $50$~K for several values of $E_i$ (130 meV, 75 meV, 25 meV from top to bottom). Each data set has been plotted subtracting the empty can background collected at the same temperature.  From the dynamic structure factor we can clearly see that there is a series of flat modes which are potentially Nd crystal field levels.  In particular, at 5 K, we detected two strong modes near 3 meV and 10 meV, and two less intense modes near 30 meV and 60 meV respectively. The level at 60 meV is quite close to an oxygen phonon ($\approx$ 58 meV) whose intensity increases in the 50 K data set. Notice that modes measured with larger incident energies are broadened in part because the energy resolution of the SEQUOIA time-of-flight spectrometer is approximately $3\%$ of the incident energy for the spectrometer settings used in these measurements.

To confirm that the modes identified above are crystal field excitations, the intensity as a function of wave-vector transfer was examined. Individual cuts as a function of $Q$ were made through the $T=5$~K data shown in Fig.~\ref{fig: cf}.  These cuts were scaled to have the same average integrated intensity for $0.5~$\AA$^{-1} < |\vec{Q}| < 3 ~$\AA$^{-1}$.  The resulting values are shown in Fig.~\ref{fig:IvsQ}.  We simultaneously fit these data to the magnetic form factor of the Nd$^{3+}$ ion (black line). The agreement is excellent validating the magnetic origin of the excitations. The Q dependence of the $\approx$ 58 meV feature has also been checked; in this case a quadratic increase of intensity as a function of Q is observed confirming its phonon nature. Indeed we found strong support for this result in Refs.~[\onlinecite{Zarembowitch,Stevie}], which showed that this phonon 
is one of the two Raman $E_g$ mode for \NO. Other phonon modes have been identified at $\approx 13$ and $\approx 23.5$ meV.

The intensity of the CEF excitations have a significant temperature dependence as shown in Fig.~\ref{fig: cf}(d-f).  These changes are due to the thermal population of each level that changes according to Boltzmann statistics.  At low temperatures, the probability that high energy levels are populated is essentially zero, so that only CEF transitions from the ground state to excited states occur as allowed by dipole selection rules. Then, as the temperature increases, the GS is depopulated in favour of the first excited state. Thermal population of the first excited state allows transitions to the other states.  Such transitions are visible at 9 and 28 meV.

\begin{figure}[t]
\includegraphics[width=0.85\columnwidth]
                {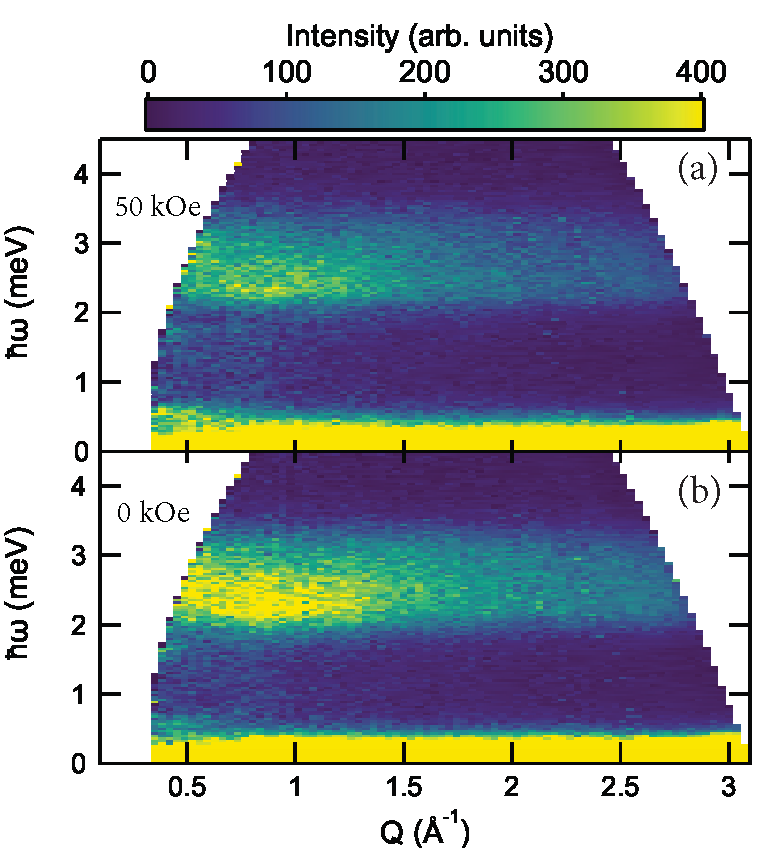}
\caption{
\label{fig: cfa}
Inelastic neutron scattering spectra of \NO~measured at $T=$5 K with $E_i=$ 6.5 meV measured with CNCS. The comparison between the (b) zero field and (a)  50 kOe applied field data shows a decrease in the intensity of the first excited state due to the Zeeman term that splits and lifts the degeneracy of the doublets. In the zero field data, there appears to be a small dispersion of the lowest excited state. }
\end{figure}

\begin{table} 
\begin{tabular}{c|cc}
\hline\hline
$B_n^m$ & LS-coupling & Intermediate\\
\hline
$B_2^0$ & $-258\pm 13$ \quad & $-281\pm 14$ \\
$B_4^0$ & $-77\pm 4$ \quad & $-71\pm 4$\\
$B_4^3$ & $56\pm 3$ \quad & $75\pm 4$ \\
$B_6^0$ & $-14.9\pm 0.7$ \quad & $-21\pm 1$ \\
$B_6^3$ & $65\pm 3$ \quad & $60\pm 3$ \\
$B_6^6$ & $57\pm 3$ \quad & $51\pm 1$ \\
\hline\hline
\end{tabular}
\caption{
\label{tab: 1}
Tabulated crystal field parameters in units of meV determined for the LS-coupling and Intermediate coupling approximations.}
\end{table}

\begin{table*} 
\begin{tabular}{c|cccccccccc}
\hline\hline
$m_J$ & $0.0$ & $0.0$ & $2.86$ & $2.86$ & $10.49$ & $10.49$ & $30.46$ & $30.46$ & $60.29$ & $60.29$\\
\hline
$-9/2$ & & & -0.099 & (-0.271) & & & & & -0.011 & (0.957)\\
$-7/2$ & 0.233 & & & & & (-0.509) & & (-0.829) & &\\
$-5/2$ & & (-0.523) & & & 0.652 & & 0.548 & & &\\
$-3/2$ & & & 0.085 & (-0.954) & & & & & -0.120 & (-0.263)\\
$-1/2$ & 0.820 & & & & & (0.561) & & (-0.114) & &\\
$1/2$ & & (0.820)& & & 0.561 & & 0.114 & & &\\
$3/2$ & & & -0.954 & (-0.085) & & & & & 0.263 & (-0.120)\\
$5/2$ & 0.523 & & & & & (-0.652) & & (0.548) & &\\
$7/2$ & & (-0.233) & & & 0.509 & & -0.829 & & &\\
$9/2$ & & & 0.271 & (-0.099) & & & & & 0.957 & (0.011)\\
\hline\hline
\end{tabular}
\caption{
\label{tab: 2}
Tabulated wave functions of the crystal field states in \NO~obtained within LS-coupling approximation. The crystal-field energies (in meV) are tabulated
horizontally, the $m_J$-values of the ground-state multiplet vertically. Only coefficients of the wave functions $> 10^{-3}$ are shown. For the sake of
representation, the wave functions of doublet excitations are gathered into one column, of which the values without (within) parentheses correspond
to the first (second) member of the doublet.}
\end{table*}

\begin{table*} 
\begin{tabular}{c|cccccccccc}
\hline\hline
$m_J$ & $0.0$ & $0.0$ & $2.86$ & $2.86$ & $10.49$ & $10.49$ & $30.46$ & $30.46$ & $60.29$ & $60.29$\\
\hline
$-9/2$ & & & -0.251 & (0.088) & & & & & 0.957 & (0.011) \\
$-7/2$ & -0.182 & & & & & (0.449) & 0.872 & & & \\
$-5/2$ & & (0.356) & & & 0.794 & & & (-0.485) & & \\
$-3/2$ & & & 0.958 & (0.086) & & & & & 0.240 & (-0.109) \\
$-1/2$ & 0.916 & & & & & (0.398) & -0.015 & & & \\
$1/2$ & & (0.916) & & &-0.398 & & & (0.015) & & \\
$3/2$ & & & -0.086 & (0.958) & & & & & -0.109 & (-0.240)\\
$5/2$ & -0.356 & & & & & (0.794) & -0.485 & & & \\
$7/2$ & & (0.182) & & & 0.449 & & & (0.872) & & \\
$9/2$ & & & 0.088 & (0.251) & & & & & -0.011 & (0.957)\\ \hline
$-11/2$ & & (-0.004) & & & 0.054 & & & (-0.028) & & \\
$-9/2$ & & &0.01 & (-0.045) & & & & & -0.070 & (0.012)\\
$-7/2$ & -0.006 & & & & & (-0.06) & -0.001 & & & \\
$-5/2$ & & (-0.015)& & & -0.033 & & & (-0.022) & & \\
$-3/2$ & & & -0.008 & (0.012) & & & & & -0.088 & (-0.029)\\
$-1/2$ & -0.008 & & & & & (-0.007) & -0.038 & & & \\
$1/2$ & & (0.008) & & & -0.007 & & & (-0.038) & & \\
$3/2$ & & & 0.012 & (0.008) & & & & & 0.029 & (-0.088)\\
$5/2$ & -0.015 & & & & & (0.033) & 0.022 & & & \\
$7/2$ & & (-0.006)& & & 0.06 & & & (0.001) & & \\
$9/2$ & & & 0.045 & (0.01) & & & & & 0.012 & (0.07)\\
$11/2$ & 0.004 & & & & & (0.054) & -0.028 & & & \\
\hline\hline
\end{tabular}
\caption{
\label{tab: 3}
Tabulated wave functions of the crystal field states in \NO~obtained within the intermediate coupling approximation, only the $J=9/2$ and $J=11/2$ mixing is presented. 
The crystal-field energies (in meV) are tabulated horizontally, the $m_J$-values of the ground-state multiplet vertically. Only coefficients of the 
wave functions $> 10^{-3}$ are shown. For the sake of representation, the wave functions of doublet excitations are gathered into one column, of which the 
values without (within) parentheses correspond to the first (second) member of the doublet.}
\end{table*}

The energy width of the first excited state can be better appreciated in Fig.~\ref{fig: cfa}, which shows the dynamic structure factor of \NO~collected at CNCS with $E_i=$ 6.5 meV at T = 5 K. The crystal field excitation is clearly broader than the instrumental resolution of 0.3 meV.  This broadening appears to be due to a weak dispersion of the crystal field excitation, likely reflecting the Nd-Nd exchange interactions. The data set collected at 50 kOe shows that the intensity of the first excited state decreases. The Zeeman term in the Hamiltonian breaks time reversal symmetry thereby lifting the degeneracy of the crystal field doublets. The additional field induced splitting of the crystal field levels alters the thermal population with the result that changes in intensity of the lowest crystal field excitations are observed.

The determination of the crystal field parameters defined in Eq.~\ref{eq: 1}, has been done using the $T=5$~K and 50 K inelastic neutron scattering data as a function of energy transfer. We fit the experimental CEF spectrum using a series of Lorentzian functions that were allowed to vary for each excitation.  We diagonalize Eq.~\ref{eq: 1} using a set of CEF parameters calculated in the point charge approximation~\cite{Hutchings,Stevens}, determine the spectrum and the eigenfunctions, and then calculate the dynamical structure factor as in Eq.~\ref{eq: 2}.  We then compare the calculation to the measurement to determine a $\chi^2$ value. This quantity is minimized by iterating the procedure until the calculation converged to a global minimum.

The crystal field parameters (in meV) determined from our refinements for both the LS-coupling and the Intermediate coupling approximations, are given in Tab.~\ref{tab: 1}. 
The extrapolated values of the ground state $g$-tensor are: $g_z = 0.231(16)$ and $g_{xy} = 1.733(12)$, with a corresponding magnetic moment of $\mu_{Nd} = 1.89(5) \mu_B$. 
The anisotropy in the $g$-tensor $g_z/g_{xy}=0.133$, where as we noted the $z$-axis is projected along the $[001]$ axis, indicates a significant XY-character of the spin.

Tables~\ref{tab: 2} and ~\ref{tab: 3} list the refined wave functions and energy levels of the crystal field states for \NO{} in the LS and intermediate coupling approximations respectively. According to our refinement, the spectrum consists of five doublets at: $0.0$, $2.86(3)$, $10.49(2)$, $30.46(3)$ and $60.29(4)$ meV in very good agreement with the experimental data. For the sake of representation, the wave functions of doublet excitations are gathered into one column, of which the values without (within) parentheses correspond to the first (second) member of the doublet. The GS is mainly a linear combination of $\psi_0=|\pm1/2\rangle$~+~$|\pm5/2\rangle$~+~$|\pm7/2\rangle$ states. This spectrum is also in excellent agreement with the energies of the crystal field excitations reported in previous optical measurements\cite{Henderson, Caro} which identified the $J=9/2$ transitions around $2.85$, $10.41$, $31.37$, $61.5$ meV and $2.60$, $9.67$, $30.25$, $60.88$ meV respectively.   


A comparison showing the fitted intensity at 5 K and 50 K versus the experimental spectra is shown in Fig.~\ref{fig: cf}(g-i). The calculation is in excellent agreement with both data sets over the entire range of energy transfer, confirming the quality of the fit.   The magnetic susceptibility calculated solely on the basis of our crystal field model is presented as a red line in Fig.~\ref{fig: susc}.


\subsection{Low Temperature Specific Heat}

\begin{figure}[t]
\includegraphics[width=0.99\columnwidth]
                {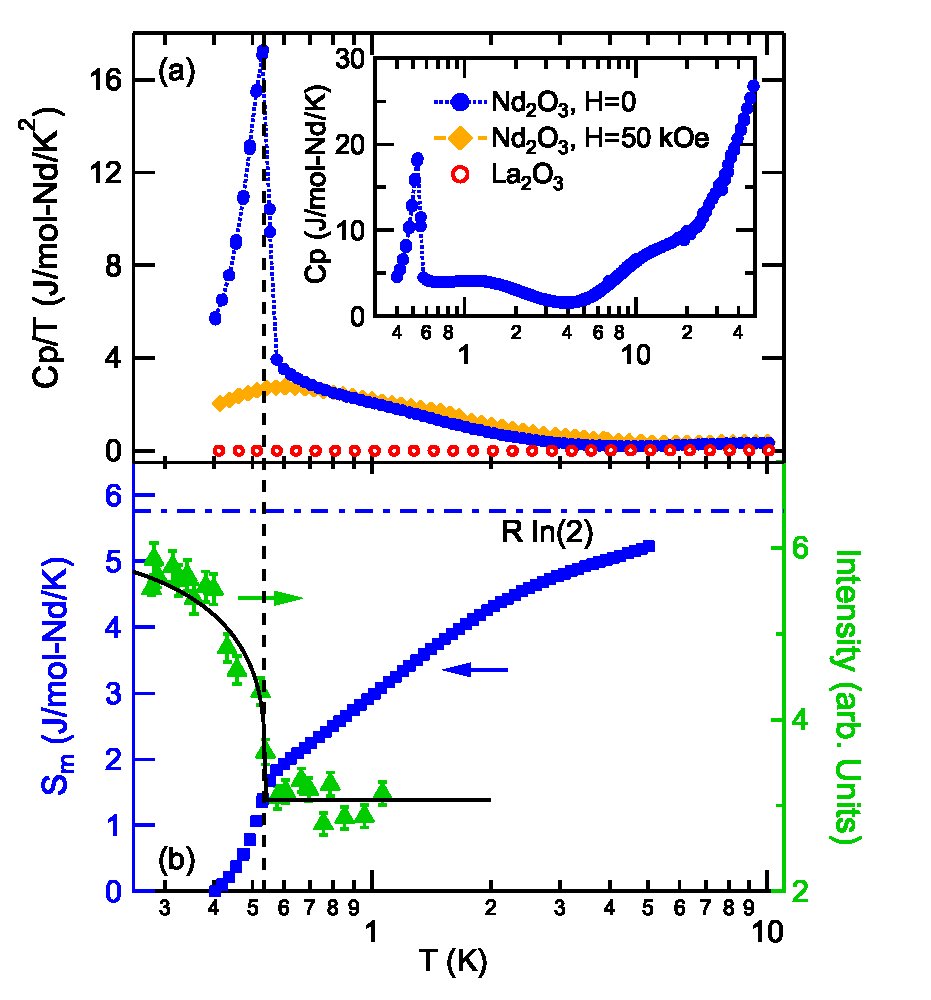}
\caption{
\label{fig: heatcap}
(a) The comparison of the Cp/T data for \NO~and La$_2$O$_3$ highlights the sharp anomaly at T=550 mK, typical of a phase transition, 
and a smaller broad contribution at 1.5 K (see inset) whose origin is at present undetermined. Both these contributions are completely absent in $\rm La_2O_3$.
A field of 50 kOe appears to either eliminate long range order or suppress it to temperatures below the observation window of the current measurements.
(b) Magnetic Bragg peak scattering intensity at (1/2 0 1/2) and magnetic entropy, $S_m$, (integrated Cp/T) from 0.4 to 5 K as a function of temperature yielding an entropy of $\approx$90\% of $R\ln(2)$. 
For this analysis the La$_2$O$_3$ was used to subtract the non magnetic background.
The vertical dashed line highlights the transition temperature of $T = 550$~mK. The solid black line is a fit of the Bragg peak scattering intensity to a power-law function and serves as a guide to the eye.
}
\end{figure}

The low temperature specific heat, $C_p/T$, is presented in Fig. \ref{fig: heatcap}(a).  The inset shows $C_p$ to higher temperatures where a Schottky anomaly is observed near 15 K.   The Schottky anomaly is expected due to the thermal depopulation of the crystal field level at 2.86 meV and is consistent with the previous work of Ref. [\onlinecite{Justice}]. Above 25 K the lattice contribution becomes an important component of the specific heat.

At temperatures below 4 K there are two previously unobserved features evident in the data.    The first is a broad contribution to the specific heat centered near 1.5 K and the second is a sharper peak near 550 mK.  As will be discussed in Sec. \ref{lro}, the low temperature peak indicates the onset of long range anti-ferromagnetic order.  A field of 50 kOe dramatically alters the specific heat and appears to either eliminate long range order or suppress it to temperatures below the observation window of the measurement.

At the present time we are unable to determine the origin of the broad feature centered near 1.5 K.  No additional signal was detected in the neutron diffraction patter at 1 K (see inset of Fig. \ref{fig: 4}). We exclude an unobserved crystal field level as an explanation since our crystal field analysis accounts for the allowed crystal field levels.  Integrating $C_p/T$ from 0.4 to 5 K yields an entropy of $\approx$90\% of $R\ln(2)$, as shown in Fig. \ref{fig: heatcap}(b). For this analysis, we utilized La$_2$O$_3$ data to subtract the nonmagnetic background.  Thus the entropy associated with the broad feature appears to be required to account for the total $R\ln(2)$ entropy expected for a doublet GS. However, this can not be stated definitively as additional entropy may be released at temperatures below those probed here.  Other possible explanations for the feature include: undetected short range magnetic order and/or a low lying magnetic excitations.  Both of these possibilities are reasonable expectations for a material with significant frustration as is the case for \NO{}. Further studies are planned to investigate this aspect of \NO{} in single crystal samples.

\subsection{Long Range Magnetic Order}\label{lro}
We use neutron diffraction measurements to probe the phase transition observed in the heat capacity measurements.  We find that \NO~has magnetic long range order below $T_N$ = 550 mK. To the best of our knowledge neither the appearance of magnetic order nor the ordered spin configuration have been previously reported for \NO.

\begin{figure}[h]
\includegraphics[width=0.99\columnwidth]
                {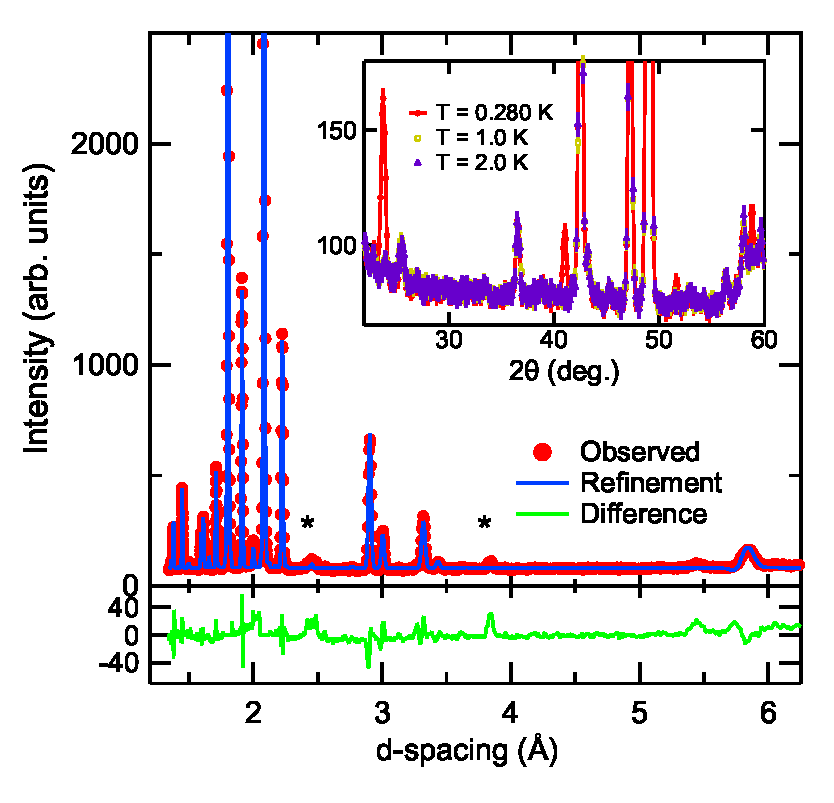}
\caption{
\label{fig: 4}
Neutron diffraction data and Rietveld refinement of the structural model for \NO{}.  Data were collected with HB-2A in zero field at $T = 280$~mK. The two strongest peaks are due to the Cu sample holder. The magnetic structure refinement used structural parameter values determined from the structural refinement at 2 K. The systems orders into an anti-ferromagnetic layered structure with spins in the basal plane, consistent with our crystal field analysis. The ordering wave vector, $\vec{k}$ = (1/2 0 1/2). The two asterisks in the main panel highlight two small impurity peaks.  The inset shows the temperature dependent diffraction data as a function of the scattering angle, $2\theta$.  Magnetic reflections are observed at $2\theta=$~ 23.8, 41.0, 51.6 and 58.9$^\circ$.  The data were collected with $\lambda$ = 2.54 \AA}
\end{figure}

Figure~\ref{fig: 4} shows the measured diffraction data and Rietveld refinement of the \NO~data collected with HB-2A in zero field at T = 280 mK. Note that two small peaks due to an unidentified impurity phase were observed in the diffraction pattern (see Fig. \ref{fig: 4}).  The structural model was refined using the 2 K data set as starting point, no evidence of a distortion or a structural phase transition was found confirming previous work performed at higher temperatures~\cite{Pauling,Faucher}. Note that the refinement was done excluding the two strongest peaks coming from the Cu holder.   The results of the structural refinements are given in Table \ref{tab: 4}.  The inset of Fig. \ref{fig: heatcap} shows the temperature dependence of the intensity of the (1/2 0 1/2) magnetic Bragg peak.  The appearance of the magnetic Bragg peak coincides with the anomaly in the specific heat at 550 mK.

The magnetic structure has been analyzed with two software packages: Full-Prof and Jana2006 to cross-check the results. In both cases this magnetic structure has been refined in the $C_c 2/m$ (\#12.63) magnetic space group with a propagation vector of (1/2 0 1/2)(referenced to the parent space group $P\bar{3}m1$)).  Within this symmetry, the Nd spins are anti-ferromagnetically coupled in the basal plane with an ordered moment of  $\mu_{Nd}$ = 1.87(10)$\mu_B$ (consistent with our crystal field analysis)(See Fig.~\ref{fig1}(b,c)).  Note that moments are constrained to be along the b-axis by the magnetic space group $C_c 2/m$ defined for a magnetic unit cell twice as large than the chemical cell along a and c directions.  Due to the close proximity of the Nd atoms to the $1/4$ position, we are unable to distinguish between the ($--++$) and ($+--+$) stacking sequence of AFM planes, as depicted in Fig. \ref{fig1}(b) and (c) respectively. The relative stacking is determined by the alternating out of  plane Nd-Nd bonds  (3.69 and 3.76 \AA).  The space group $C_c 2/m$ allows both solutions and 
the calculated structure factor does not change appreciably between the two models.  The results of the refinement of the magnetic structure are given in Table \ref{tab: 4}. 

\begin{table}[t] 
\begin{tabular}{c|lc}
\hline\hline
Temperature & \qquad \qquad \quad Position & \quad GOF \\
\hline
2 K & a=b=3.83(2) \AA,~c=6.00(3) \AA & \quad 1.70 \\
& Nd = 1/3, 2/3, 0.246(3) & \\
& O(1) = 0, 0, 0 & \\
& O(2) = 1/3, 2/3, 0.648(5) & \\
\hline
280 mK & a=b=3.83 \AA,~c=6.00 \AA & \quad 1.91\\
& Nd = 1/3, 2/3, 0.246(4) & \\
& O(1) = 0, 0, 0 & \\
& O(2) = 1/3, 2/3, 0.647(7) & \\
& $\mu_{Nd}$ = 1.87(10)$\mu_B$ \\
\hline\hline
\end{tabular}
\caption{
\label{tab: 4}
Rietveld refinement summary for \NO~ in the space group $P\bar{3}m1$ (164).
Notice that the refinement at 280 mK has been done keeping the 2 K lattice constants fixed.}
\end{table}

Taken together the results above indicate that \NO{} is a chemically and structurally simple frustrated system. Since the ground state eigenfuction is mostly composed of $J_z$=1/2 (67 \%) and is reasonably well separated from the first excited state crystal field level at 2.86 meV it is temping to view \NO{} has an effective spin 1/2 system with potentially strong symmetric anisotropic exchange terms due to the strong spin-orbit coupling expected for a rare-earth ion.  

It is also interesting that the stripe like magnetic order of \NO{} is similar to that found in theoretical models of triangular lattices.  This type of stripe order is nearby to a spin liquid phase\cite{zhu_spin_liquid}.  However, the situation in \NO{} appears to be rather different from a geometrically frustrated triangular lattice.  As described above, in \NO{} the nearest neighbor Nd-Nd distances are out of plane.   Thus the frustration appears to be driven by competing interactions rather than geometrical frustration.  A definitive answer to the importance of various exchange parameters awaits studies of single crystal samples.





\section{Conclusions}

We have examined the magnetic properties and crystal field levels of the binary compound \NO. The crystal field spectrum spectrum of the Nd$^{3+}$ ions spans an energy range between 3 and 60 meV; in zero field the XY-like spins are in the ab plane with a magnetic moment of 1.89(5) $\mu_B$/ion. This value is consistent with the ordered moment of 1.87(10)$\mu_B$ determined from the neutron diffraction data. The ordered phase consists of spins AFM coupled in the ab plane, and oriented along b-axis with an ordering vector $\vec{k}$ = (1/2,0,1/2).  The planar arrangement of ordered moments agrees with the XY anisotropy determined from the crystal field levels. The results presented here suggest that \NO{} is a strongly frustrated system driven by competing interactions.


\begin{acknowledgments}
We acknowledge useful discussions with C. Batista.  This work was supported by the U.S. DOE, Office of Science, Basic Energy Sciences, Materials Sciences and Engineering Division. This research used resources at the High Flux Isotope Reactor and Spallation Neutron Source, DOE Office of Science User Facilities operated by the Oak Ridge National Laboratory. HSA and DM acknowledge support from the Gordon and Betty Moore Foundation’s EPIQS Initiative through Grant GBMF4416.
\end{acknowledgments}


\end{document}